\documentclass[prb,twocolumn,superscriptaddress,tightenlines,longbibliography]{revtex4-2}

\usepackage[margin=1in]{geometry}
\usepackage{amsmath,amssymb,amsfonts,mathtools}
\usepackage{graphicx}
\usepackage{xcolor}
\usepackage{tikz}
\usetikzlibrary{arrows.meta,decorations.markings,calc,positioning,decorations.pathreplacing,fit,backgrounds}
\usepackage{pgfplots}
\pgfplotsset{compat=1.18}
\usepgfplotslibrary{fillbetween}
\usepackage{booktabs}
\usepackage{array}
\usepackage{enumitem}
\usepackage{hyperref}
\usepackage{cleveref}
\usepackage{microtype}
\usepackage{float}

\hypersetup{
  colorlinks=true,
  linkcolor=blue!50!black,
  citecolor=blue!50!black,
  urlcolor=blue!50!black
}
\definecolor{Zgreen}{RGB}{178,223,138}
\definecolor{Xpink}{RGB}{251,180,185}
\definecolor{pairorange}{RGB}{230,126,34}
\definecolor{softgray}{RGB}{150,150,150}
\definecolor{darkgray}{RGB}{80,80,80}
\definecolor{bandgray}{RGB}{238,238,238}

\definecolor{colA}{RGB}{190,55,45}
\definecolor{colB}{RGB}{42,92,170}
\definecolor{colC}{RGB}{20,135,105}
\definecolor{colD}{RGB}{205,125,25}
\definecolor{xfill}{RGB}{247,205,207}
\definecolor{zfill}{RGB}{207,235,205}
\definecolor{datafill}{RGB}{246,246,246}

\newtheorem{proposition}{Proposition}
\newtheorem{definition}{Definition}

\newcommand{\Ftwo}{\mathbb F_2}

\newcommand{\wt}{\operatorname{wt}}
\newcommand{\im}{\operatorname{im}}

\newcommand{\Z}{\mathbb Z}

\newcommand{\Lam}{\Lambda}
\newcommand{\Lamp}{\Lambda'}

\newcommand{\F}{\mathbb F_2}

\newcommand{\R}{\mathcal R}
\newcommand{\supp}{\operatorname{supp}}

\newcommand{\Hz}{H_Z^{\mathrm{pair}}}
\newcommand{\Hx}{H_X^{\mathrm{pair}}}

\newcommand{\cJ}{\mathcal J}
\newcommand{\cP}{\mathcal P}
\newcommand{\cD}{\mathcal D}

\newcommand{\Nminus}[2]{\mathcal N^{-}_{#1}(#2)}
\newcommand{\Nplus}[2]{\mathcal N^{+}_{#1}(#2)}

\begin{document}
\title{Floquet Abelian Multicycle Codes}

\author{Alexey A. Kovalev}

\affiliation{Department of Physics and Astronomy and Nebraska Center for Materials and Nanoscience, University of Nebraska, Lincoln, Nebraska 68588, USA}

\date{\today}

\begin{abstract}
Abelian multicycle (AMC) codes are compact quantum
low-density parity-check codes whose multiblock
chain-complex structure provides redundant low-weight
stabilizers and supports single-shot error correction.
We introduce Floquet AMC codes by deriving a
quotient-lattice representation of a general level-$j$,
$D$-dimensional AMC complex over a finite Abelian group
algebra, lifting this lattice to spacetime, and rotating the
circuit-time direction in the associated ZX network.
When the check and data spiders have even valence and admit
a time-oriented local port matching, the network decomposes
into a periodic schedule of native two-qubit $XX$ and $ZZ$
measurements. We construct generalized-bicycle and level-$2$
AMC4 examples, determine their instantaneous stabilizer
groups, and compute their embedded distances by minimizing
over all inequivalent circuit cuts. For AMC4 instances
locally equivalent to four-dimensional toric codes, we
obtain Floquet memories with parameters
$[[108,6,5]]$, $[[144,6,8]]$, and $[[324,6,10]]$.
Local Pauli-web detector templates and beam-search decoding
under the measurement-native EM3 noise model yield an
estimated pseudothreshold of approximately $1.2\%$.
These results provide compact measurement-only realizations
of higher-dimensional homological redundancy without
directly measuring the original weight-six stabilizers.
\end{abstract}
\maketitle

\section{Introduction}

Surface and toric codes remain the canonical examples of quantum memories because of their locality and high thresholds \cite{Kitaev2003,Dennis2002,Preskill1998,Terhal2015}, but their two-dimensional locality also imposes strong tradeoffs among block length $n$, encoded dimension $k$, and distance $d$ \cite{BravyiPoulinTerhal2010}. This has motivated the search for quantum low-density parity-check (qLDPC) codes~\cite{PRXQuantum.2.040101} with better finite-size parameters and simplified syndrome-extraction circuits.
Two-block Kronecker sum-product codes constructed from circulant matrices (Sec.~III C of Ref.~\cite{KovalevPryadko2013}) are equivalent to two-block group-algebra (2BGA) codes and, in simple cases, lead to generalized bicycle codes~\cite{KovalevPryadko2013,MacKayMitchisonMcFadden2004}. Two-block Kronecker sum-product and 2BGA constructions yield compact CSS codes~\cite{KovalevPryadko2013,WangPryadko2022,WangLinPryadko2023,LinPryadko2024}; bivariate bicycle codes are a prominent subfamily with exceptional finite-size performance~\cite{BravyiBB2024}.
In a fault-tolerant setting, however, stabilizer codes are not fully characterized by the stabilizer group alone. The full spacetime syndrome-extraction process includes measurement faults, hook errors, and redundant checks, and its performance can be limited by the decoder’s ability to use syndrome information with low latency.

Single-shot schemes are particularly useful in this setting because the redundancy present in each measured batch of stabilizers simplifies error correction~\cite{Bombin2015,BrownNickersonBrowne2016,Campbell2019}.  Higher-dimensional toric and hypergraph-product codes~\cite{TillichZemor2014} and their variations~\cite{KovalevPryadko2012ImprovedHGP} provide a natural homological source of such redundancy and can support single-shot or few-shot decoding~\cite{ZengPryadko2019,ZengPryadko2020,Quintavalle2021,HiggottBreuckmann2023}.  Abelian multicycle (AMC) codes were introduced to retain these higher-dimensional features in much shorter block lengths.  They are built from multiblock chain complexes over abelian group algebras, generalizing the relation between two-block codes and hypergraph-product codes to arbitrary complex dimension \cite{LinAMC2026}.  In the important four-dimensional examples, AMC codes are locally equivalent to rotated four-dimensional toric codes, have low-weight redundant stabilizer generators, and exhibit single-shot decoding~\cite{AasenGeometric2025,AasenTopological2025,LinAMC2026}.

Floquet codes use a periodic sequence of generally noncommuting measurements to generate and maintain logical qubits; the instantaneous stabilizer group (ISG) changes during the cycle, and deterministic information is recovered from spacetime detectors~\cite{HastingsHaah2021,DavydovaTantivasadakarnBalasubramanian2023,FuGottesman2025}.  The honeycomb code showed that such a memory can be built entirely from two-qubit Pauli measurements \cite{HastingsHaah2021,Gidney2021}, and later hyperbolic and semi-hyperbolic Floquet codes improved the encoding rate while preserving a measurement-native structure \cite{HiggottBreuckmann2024,Fahimniya2025}.  Starting with a static qLDPC code and performing a direct replacement of each stabilizer measurement by pairwise measurement gadgets may introduce many ancillas, long idling times, and distance-reducing correlated faults \cite{Chao2020,Gidney2023,Rodatz2024}.  An alternative approach was demonstrated by the Stairway construction of Floquet codes, which applies the ZX calculus to 2BGA codes~\cite{Bombin2024,Backens2014,Wetering2020,Stairway2026}.  This converts the static algebraic code into a Floquet memory whose design freedom is largely shifted to the choice of periodic boundary conditions and local worldline pairings~\cite{Stairway2026}.

In this work, we introduce \emph{Floquet AMC codes}, a measurement-driven extension of AMC codes that combines the redundant homological structure of multiblock group-algebra complexes with the pairwise-measurement framework of Floquet codes.  The starting point is an AMC complex specified by commuting group-algebra elements $a_1,\ldots,a_D\in\mathbb{F}_2[G]$ and a homological level $j$.  We represent the associated syndrome-extraction process as a foliated ZX network, choose a time covector that is not aligned with the original circuit time direction, and reinterpret selected edges as physical qubit worldlines.  When the relevant ZX nodes have even degree, the network admits a decomposition into a periodic sequence of two-qubit Pauli measurements.  The resulting protocol is dynamical: stabilizers need not persist from one substep to the next, while detector relations arise from the chain-complex identities and in some cases from local redundancies introduced by the spider decompositions. A Floquet version of an AMC code can therefore encode this redundancy in a spacetime detector structure rather than as a static list of checks that must all be measured in each round.  This offers a route to measurement-native qLDPC memories that retain the compact block lengths of group-algebra codes while inheriting part of the confinement and few-shot decoding behavior of higher-dimensional homological codes.

The remainder of this article is organized as follows. Section~II reviews AMC complexes, the associated quantum codes, and the relevant ZX-calculus representation, and it summarizes the procedure for identifying the instantaneous stabilizer group (ISG) and logical operators. Section~III constructs the quotient lattice for a general AMC code. This lattice construction naturally supports the Floquetification described in Section~IV, where we introduce the rotated time direction, qubit worldlines, and pairwise-measurement factorization. Section~V presents Floquet constructions for two- and four-dimensional AMC codes. Section~VI reports numerical results for four-dimensional AMC codes and estimates a pseudothreshold of approximately $1.2\%$. Section~VII concludes the article.

\begin{figure}
\centering
\includegraphics{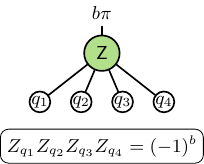}
\quad
\includegraphics{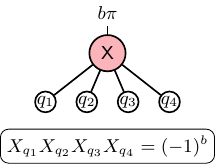}
\caption{Spiders as check measurements for the case with no output states. Here $b$ is an integer that describes the measurement parity. The number of qubits is arbitrary and depends on the particular stabilizer code. The color indicates whether the measured parity is $Z$-type or $X$-type.}
\label{fig:spiders-as-checks}
\end{figure}
\section{Methods}
\subsection{ZX-calculus construction}
The ZX calculus provides a graphical language for quantum
processes and interacting complementary
observables~\cite{CoeckeDuncan2011,Backens2014}.
A recent ZX-calculus construction starts from a static CSS code~\cite{CalderbankShor1996,Steane1996}, represents its syndrome-extraction circuit as a ZX tensor network, rotates the time direction, and decomposes high-degree spiders into pairwise measurements~\cite{Stairway2026}. The $Z$ and $X$ spiders can be introduced as objects with $n$ input states, $m$ output states, and a phase $\alpha$. In the operator notation, the $X$ spider is defined as
\begin{equation}
|+\rangle^{\otimes m}\langle +|^{\otimes n}
+
e^{i\alpha}
|-\rangle^{\otimes m}\langle -|^{\otimes n},
\end{equation}
and the $Z$ spider is defined as
\begin{equation}
|0\rangle^{\otimes m}\langle 0|^{\otimes n}
+
e^{i\alpha}
|1\rangle^{\otimes m}\langle 1|^{\otimes n}.
\end{equation}
In the absence of output states, this construction describes a Pauli measurement as a projection associated with the measurement of a $Z$ or $X$ check, with $\alpha=b\pi$, as shown in Fig.~\ref{fig:spiders-as-checks}. 

\begin{figure}
\centering
\includegraphics{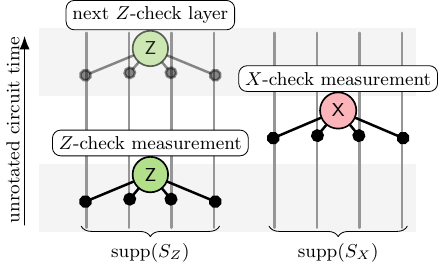}
\caption{Unrotated check-measurement network.  Data-qubit worldlines run vertically through alternating check-measurement layers. The unrotated network can be viewed as a syndrome-extraction circuit drawn in ZX notation.}
\label{fig:unrotated-check-measurement}
\end{figure}

We first consider the ordinary syndrome-extraction network, drawn in the original circuit-time direction, in which the data-qubit worldlines run vertically; see Fig.~\ref{fig:unrotated-check-measurement}.
A $Z$-check measurement layer couples a $Z$-spider to the data qubits in the support of a $Z$ stabilizer, and an $X$-check measurement layer similarly couples an $X$-spider to the support of an $X$ stabilizer.
The syndrome-extraction process can then be represented as a ZX network, as shown in Fig.~\ref{fig:unrotated-check-measurement}.
This network can be reinterpreted as a measurement-only Floquet schedule by assigning worldlines, rotating the time axis, and decomposing the spiders into pairwise measurements. 

\begin{figure}
\centering
\includegraphics{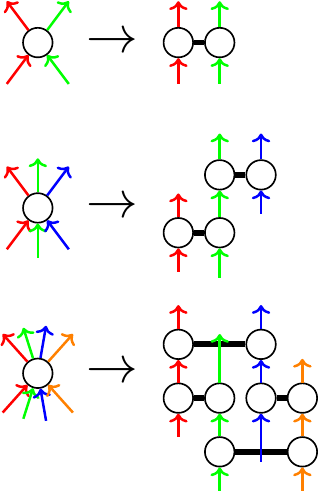}
\caption{Stairway-style decomposition of many-leg spiders into pairwise measurements. The measured Pauli type is inherited from the spider color, pink (X) or green (Z).}
\label{fig:stairway-style-decomposition}
\end{figure}

For the last step, we use the fact that an even-legged spider can be decomposed into native two-qubit parity measurements, as shown in Fig.~\ref{fig:stairway-style-decomposition}~\cite{Stairway2026}.
For a $2n$-leg spider interpreted as an $n\to n$ operator, a matching pairs the incoming and outgoing legs.
Each matched pair is then measured with a two-qubit parity measurement of the same Pauli type.
The resulting gadget implements the same external projector for a fixed outcome configuration, and other outcomes are absorbed into a Pauli frame.
For eight-legged spiders, one may include one redundant pairwise measurement; this introduces a local detector~\cite{Stairway2026}. The check network first determines which many-legged spiders are present, while the matching specifies how to pair the legs of each spider and thereby determines the pairwise-measurement schedule.
The induced stabilizer group, detectors, and logical operators are properties of this final pairwise schedule.

\subsection{Abelian two-block group-algebra codes}
\label{sec:2bga-definition}

Let \(G\) be a finite Abelian group of order
\(\ell:=|G|\), and let
\[
\mathbb F_2[G]
=
\left\{
a=\sum_{g\in G}a_g g:
a_g\in\mathbb F_2
\right\}
\]
denote its binary group algebra.  For each \(g\in G\), let
\(\rho(g)\in\operatorname{Mat}_{\ell}(\mathbb F_2)\) be the
left-regular permutation matrix,
\begin{equation}
 [\rho(g)]_{h,k}
 =
 \begin{cases}
  1, & h=gk,\\
  0, & \text{otherwise},
 \end{cases}
 \qquad h,k\in G.
\end{equation}
The regular representation extends linearly to
\(
\rho(a)
=
\sum_{g\in G}a_g\rho(g)
\).
We choose two group-algebra elements
\begin{equation}
a=\sum_{g\in G}a_g g,
\qquad
b=\sum_{g\in G}b_g g,
\end{equation}
and define the corresponding binary matrices
\begin{equation}
A:=\rho(a),
\qquad
B:=\rho(b).
\end{equation}
The associated Abelian two-block group-algebra
(\(2\)BGA) code is the CSS stabilizer code~\cite{CalderbankShor1996,Steane1996} with check
matrices~\cite{WangPryadko2022}
\begin{equation}
 H_X=
 \begin{bmatrix}
  A | B
 \end{bmatrix},
 \qquad
 H_Z=
 \begin{bmatrix}
  B^{\mathsf T} | A^{\mathsf T}
 \end{bmatrix}.
\label{eq:2bga1}
\end{equation}
Since \(G\) is Abelian, the matrices \(A\) and \(B\) commute.

The code contains
\(
 n=2|G|=2\ell
\)
physical qubits, with one qubit of each block type associated
with every \(g\in G\).  
When \(G\) is cyclic, the construction reduces to a
generalized bicycle code~\cite{KovalevPryadko2013}, while a product of two cyclic
groups gives the bivariate bicycle family~\cite{BravyiBB2024}. 

\subsection{Two-block Kronecker sum-product codes from circulant matrices}
Two-block Kronecker sum-product codes were defined in Sec.~III C of Ref.~\cite{KovalevPryadko2013} using Eq.~\eqref{eq:2bga1}, with matrices
\begin{align}
 A
 &=
 \sum_{\boldsymbol{i}\in\mathcal I_A}
 H^A_{i_1,1}\otimes\cdots\otimes H^A_{i_m,m},
 \nonumber\\
 B
 &=
 \sum_{\boldsymbol{i}\in\mathcal I_B}
 H^B_{i_1,1}\otimes\cdots\otimes H^B_{i_m,m},
 \label{eq:tp}
\end{align}
where
\(
\boldsymbol{i}=(i_1,\ldots,i_m),
\)
and circulant matrices
\begin{equation}
 H^K_{i_j,j}=f^K_{i_j,j}(P_j),
 \quad
 f^K_{i_j,j}(t)
 \in
 \mathbb F_2[t]/(t^{L_j}-1),
\end{equation}
for $K\in\{A,B\}$.
Here $P_j$ is the $L_j\times L_j$ basic cyclic-shift
matrix associated with the cyclic group $C_{L_j}$.
Since all matrices appearing in the $j$th tensor factor
are polynomials in the same matrix $P_j$, all terms in
$A$ and $B$ commute, and hence $AB=BA$.

Let
\[
G=C_{L_1}\times\cdots\times C_{L_m},
\]
and denote by $x_j$ a generator of the $j$th cyclic
factor. There is a natural algebra isomorphism
\begin{equation}
 \mathbb F_2[G]
 \cong
 \frac{\mathbb F_2[x_1,\ldots,x_m]}
 {(x_1^{L_1}-1,\ldots,x_m^{L_m}-1)}.
\label{eq:product-group-algebra}
\end{equation}
With the group elements ordered lexicographically, the
regular representation satisfies
\begin{equation}
 \rho_G(x_j)
 =
 I_{L_1}\otimes\cdots\otimes I_{L_{j-1}}
 \otimes P_j
 \otimes I_{L_{j+1}}\otimes\cdots\otimes I_{L_m}.
\end{equation}
Consequently, the matrices in Eq.~\eqref{eq:tp} are the
regular-representation matrices
\begin{equation}
 A=\rho_G(a),
 \qquad
 B=\rho_G(b),
\end{equation}
of the group-algebra elements
\begin{align}
 a
 &=
 \sum_{\boldsymbol{i}\in\mathcal I_A}
 \prod_{j=1}^{m}f^A_{i_j,j}(x_j),
 \label{eq:tensor-a}\\
 b
 &=
 \sum_{\boldsymbol{i}\in\mathcal I_B}
 \prod_{j=1}^{m}f^B_{i_j,j}(x_j).
 \label{eq:tensor-b}
\end{align}
Thus every two-block Kronecker sum-product code of the form given in
Eqs.~\eqref{eq:2bga1} and \eqref{eq:tp} is an Abelian 2BGA code over
$\mathbb F_2$.

Conversely, every finite Abelian group is isomorphic to
a finite direct product of cyclic groups. Therefore, after
a permutation of the group-coordinate basis, every Abelian
2BGA code over $\mathbb F_2$ can be represented in the
two-block Kronecker sum-product form of Eq.~\eqref{eq:tp}.
Following Ref.~\cite{LinAMC2026}, every finite Abelian group admits a presentation
\begin{equation}
 G\cong G_\Delta
 :=
 \mathbb Z^m/\Delta\mathbb Z^m,
\label{eq:G-delta}
\end{equation}
where $\Delta\in\operatorname{Mat}_m(\mathbb Z)$ is the integer relations matrix and it has
full rank. Suppose that the two 2BGA blocks are
\begin{equation}
 A'=\rho_\Delta(a'),
 \quad
 B'=\rho_\Delta(b'),
 \quad
 a',b'\in\mathbb F_2[G_\Delta].
\end{equation}
For a more convenient representation on a lattice, let
\begin{equation}
 U\Delta V
 =
 S
 =
 \operatorname{diag}(s_1,\ldots,s_m),
 \quad
 U,V\in\operatorname{GL}_m(\mathbb Z),
\label{eq:delta-snf}
\end{equation}
be the Smith normal form of $\Delta$ with \(
1\leq s_1\mid s_2\mid\cdots\mid s_m
\).
Then
\begin{equation}
 G_\Delta
 \cong
 G_S
 :=
 C_{s_1}\times\cdots\times C_{s_m},
\label{eq:group-snf-decomposition}
\end{equation}
where factors with $s_i=1$ may be omitted.

For the column-lattice convention in Eq.~\eqref{eq:G-delta},
the induced isomorphism may be written as
\begin{equation}
 \phi:G_\Delta\longrightarrow G_S,
 \qquad
 \phi\bigl([v]_\Delta\bigr)=[Uv]_S.
\end{equation}
Extending $\phi$ linearly gives an isomorphism of group
algebras
\(
\phi:\mathbb F_2[G_\Delta]
\longrightarrow
\mathbb F_2[G_S]
\).
Let $\Pi_\phi$ be the permutation matrix defined by
\begin{equation}
 \Pi_\phi e_g=e_{\phi(g)},
 \qquad g\in G_\Delta.
\end{equation}
Then, for every $c\in\mathbb F_2[G_\Delta]$,
\begin{equation}
 \Pi_\phi\rho_\Delta(c)\Pi_\phi^{-1}
 =
 \rho_S\bigl(\phi(c)\bigr).
\label{eq:regular-representation-intertwiner}
\end{equation}
In particular,
\begin{align}
 A_S
 &:=
 \Pi_\phi A'\Pi_\phi^{-1}
 =
 \rho_S\bigl(\phi(a')\bigr),
 \\
 B_S
 &:=
 \Pi_\phi B'\Pi_\phi^{-1}
 =
 \rho_S\bigl(\phi(b')\bigr).
\end{align}
Since $G_S$ is a product of cyclic groups, $A_S$ and
$B_S$ are sums of Kronecker products of circulant
matrices and therefore have the form in Eq.~\eqref{eq:tp}.
The Smith-normal-form transformation corresponds
to a relabeling of the qubits, together
with a relabeling of the $X$- and $Z$-check generators.

\subsection{Abelian multicycle codes}
A \(D\)-dimensional AMC complex can be defined from the group-algebra elements $a_1,\ldots,a_D$ or from the corresponding commuting matrices $A_1,\ldots,A_D$~\cite{LinAMC2026}. The notation
\begin{equation}
\mathcal Q^{(D)}
=
\operatorname{AMC}(a_1,\ldots,a_D)
\label{eq:complex}
\end{equation}
defines the chain complex
\begin{equation}
0
\leftarrow
\mathbb F_2^{\binom{D}{0}\ell}
\xleftarrow{\;Q_1^{(D)}\;}
\mathbb F_2^{\binom{D}{1}\ell}
\xleftarrow{\;Q_2^{(D)}\;}
\cdots
\xleftarrow{\;Q_D^{(D)}\;}
\mathbb F_2^{\binom{D}{D}\ell}
\leftarrow
0,
\end{equation}
where $Q_j^{(D)}
\in
\operatorname{Mat}_{\binom{D}{j-1}\ell
\times
\binom{D}{j}\ell}(\mathbb F_2)$ are the boundary matrices.

The construction can be defined recursively, starting from the Abelian 2BGA case, which can be regarded as a two-dimensional chain complex:
\begin{equation}
Q_1^{(2)}
=
\begin{bmatrix}
A_2 | A_1
\end{bmatrix},
\qquad
Q_2^{(2)}
=
\begin{bmatrix}
A_1\\
A_2
\end{bmatrix}.
\end{equation}
Suppose $\mathcal Q^{(D)}$ has been constructed from $a_1,\ldots,a_D$.
Given an additional group-algebra element $a_{D+1}$ and the corresponding matrix representation $A_{D+1}$, we define the chain complex
\begin{equation}
\mathcal Q^{(D+1)}
=
\operatorname{AMC}(a_1,\ldots,a_D,a_{D+1}),
\end{equation}
by the boundary matrices
\begin{equation}
Q_1^{(D+1)}
=
\begin{bmatrix}
A_{D+1} | Q_1^{(D)}
\end{bmatrix},
\end{equation}
\begin{equation}
Q_i^{(D+1)}
=
\begin{bmatrix}
Q_{i-1}^{(D)} & 0\\
I_{\binom{D}{i-1}}\otimes A_{D+1} & Q_i^{(D)}
\end{bmatrix},
\quad
2\leq i\leq D,
\end{equation}
and
\begin{equation}
Q_{D+1}^{(D+1)}
=
\begin{bmatrix}
Q_D^{(D)}\\
A_{D+1}
\end{bmatrix}.
\end{equation}

Since all blocks \(A_i\) commute, the recursive boundary
matrices satisfy
\(
Q_i^{(D)} Q_{i+1}^{(D)}=0
\), for all $1\leq i<D$.
Therefore, we can define the level-\(j\) AMC CSS code using
the stabilizer check matrices
\begin{equation}
H_X
=
Q_j^{(D)},
\qquad
H_Z
=
\left(Q_{j+1}^{(D)}\right)^T.
\end{equation}
The number of physical qubits is
\begin{equation}
n_j
=
\binom{D}{j}|G|
=
\binom{D}{j}\ell,
\end{equation}
and the number of encoded qubits is
\(
k_j
=
n_j
-
\operatorname{rank} Q_j^{(D)}
-
\operatorname{rank} Q_{j+1}^{(D)}
\).

\subsection{Instantaneous stabilizer groups for Floquet codes}
The Bell-reference method~\cite{Stairway2026} extracts the ISG of floquet code, which can also be expressed via the stabilizer formalism~\cite{FuGottesman2025}. Let $Q$ be the physical register of $n$ qubits and let $R$ be a reference register of $n$ qubits. Initialize the joint state as
\begin{equation}
|\Psi\rangle_{QR}=\bigotimes_{i=1}^n |\phi^+\rangle_{Q_iR_i}.
\end{equation}
The measurement schedule is applied only to $Q$, and the final stabilizer tableau is represented in binary symplectic form with columns
\begin{equation}
(X_Q\mid Z_Q\mid X_R\mid Z_R).
\end{equation}
After a measurement cycle is completed, the ISG is obtained by taking row combinations with zero support on $R$ after the first Gaussian-elimination pass. A second elimination computes a basis for the normalizer quotient
\(
\mathcal N(\mathrm{ISG})/\mathrm{ISG},
\)
which is the non-gauge logical space.

For constructions that begin with a static CSS code, the ISG admits a CSS generator representation.

\section{Static codes and quotient lattice}
\subsection{2BGA codes}

We consider an Abelian 2BGA CSS code constructed from two group-algebra elements $a$ and $b$; see Eq.~\eqref{eq:2bga1}. Such a code corresponds to the two-dimensional case of the AMC construction. Under the identification $A_2=A$, $A_1=B$,
the level-$1$ matrix $Q_1^{(2)}$ reproduces the
$2$BGA convention in Eq.~\eqref{eq:2bga1}.
After multiplying $a$ and $b$ by group elements, which only permutes rows and columns, one may assume that the identity $e\in G$ occurs in both supports.  We can write
\begin{align}
 \supp(a)&=\{e,g_{A,1},\ldots,g_{A,r_A}\},\\
 \supp(b)&=\{e,g_{B,1},\ldots,g_{B,r_B}\},
\end{align}
and introduce the lattice
\(
 L=\Z^{d},
\) with $d=r_A+r_B$
and basis-vector families~\cite{Arnault2026}
\begin{equation}
 \cJ_A=\{\mathbf j_{A,1},\ldots,\mathbf j_{A,r_A}\},\quad
 \cJ_B=\{\mathbf j_{B,1},\ldots,\mathbf j_{B,r_B}\}.
\end{equation}
We further define the group homomorphism
\begin{equation}
 \Phi: L\longrightarrow G,
 \quad
 \Phi(\mathbf j_{A,\mu})=g_{A,\mu},\quad
 \Phi(\mathbf j_{B,\nu})=g_{B,\nu},
 \label{eq:phi-2bga}
\end{equation}
and let
\(
 \Lambda=\ker\Phi
\).
If the support elements generate $G$, then
\(
 L/\Lambda\simeq G
\).
If they generate only a subgroup $G_0<G$, the Tanner graph decomposes into $[G:G_0]$ isomorphic connected components, and the same construction applies with $G_0$ in place of $G$.

At every vertex $u\in L/\Lambda$, we place two qubit slots, $q_A(u)$ and $q_B(u)$, and one check of each CSS type, $X(u)$ and $Z(u)$~\cite{Arnault2026}; see Fig.~\ref{fig:matrix-exact-incidence} for an example.  The exact local incidences are
\begin{align}
 X(u):\quad& q_A(u),\, \{q_A(u-\mathbf j):\mathbf j\in\cJ_A\},\nonumber \\
                & q_B(u),\ \{q_B(u-\mathbf j):\mathbf j\in\cJ_B\},
 \label{eq:2bga-X-incidence}
 \end{align}
for the $X$ checks. For the $Z$ checks, we obtain
 \begin{align}
 Z(u):\quad& q_A(u),\, \{q_A(u+\mathbf j):\mathbf j\in\cJ_B\},\nonumber \\
              & q_B(u),\ \{q_B(u+\mathbf j):\mathbf j\in\cJ_A\}.
 \label{eq:2bga-Z-incidence}
\end{align}
Thus an $X$ row sees the identity term at the cell itself and the nonidentity terms in negative lattice directions.  Transposition reverses the translations, so a $Z$ row sees the corresponding positive directions; the two block types are exchanged because of the order $[B^{\mathsf T}\mid A^{\mathsf T}]$.

\begin{proposition}~\cite{Arnault2026}
Assume that the support elements used to define
$\Phi: L\to G$ generate $G$, and let
$\Lambda=\ker\Phi$. Then the bipartite graph defined by Eq.~\eqref{eq:2bga-X-incidence} and \eqref{eq:2bga-Z-incidence} on $L/\Lambda$ is isomorphic, up to row and column permutations, to the Tanner graph of the 2BGA code.
\end{proposition}

\emph{Reason.}  Label a quotient vertex $u+\Lambda$ by $\Phi(u)\in G$.  Translation by $-\mathbf j_{A,\mu}$ changes the label by $g_{A,\mu}^{-1}$, exactly as in a row of $\rho(g_{A,\mu})$; translation by $+\mathbf j_{A,\mu}$ gives the transpose incidence.  The same statement holds for the $B$ family.  Since two points are identified precisely when their difference lies in $\ker\Phi$, no additional or missing incidences occur.

The resulting $L$ has dimension
\begin{equation}
 d=(\wt a-1)+(\wt b-1)=w-2,
\end{equation}
where $w=\wt a+\wt b$ is the stabilizer weight~\cite{Stairway2026,Arnault2026}.

The Floquetification procedure uses this unit-cell representation by splitting each unit cell into $X$- and $Z$-type half-cells that share the same data qubits; see Fig.~\ref{fig:matrix-exact-incidence}. The data qubits and checks are then represented by $X$ or $Z$ spiders, and a Floquet worldline assignment is attached to every spider, see details in the following section. 

\subsection{AMC4 codes}
Let $G$ be a finite Abelian group with group algebra $\mathcal R=\F[G]$. We consider four nonzero group-algebra elements
\begin{equation}
 a_J=\sum_{g\in G}a_{J,g}g\in\mathcal R,
 \quad J\in\cD:=\{A,B,C,D\},
 \label{eq:four-group-algebra-elements}
\end{equation}
and represent them by the regular-action matrices
\begin{equation}
 A=\rho(a_A), B=\rho(a_B), C=\rho(a_C), D=\rho(a_D).
 \label{eq:four-regular-matrices}
\end{equation}
Because $G$ is Abelian, these four matrices commute, which is the algebraic condition needed in the AMC chain complex.  
\begin{figure*}[t]
\centering
\includegraphics{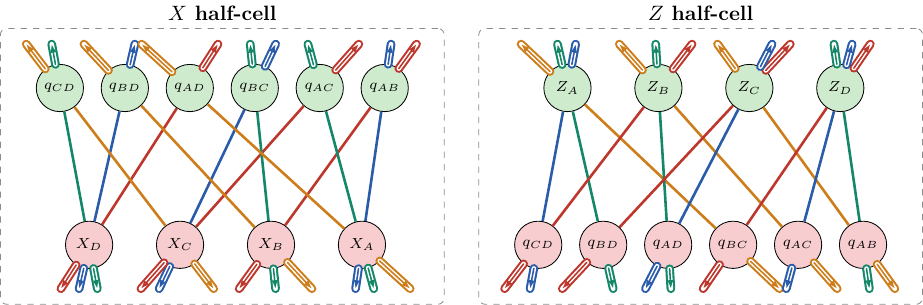}
\caption{Intracell edges and external connection bundles for the AMC4 half-cells over an arbitrary finite Abelian group algebra. The thin colored lines inside each dashed box are the identity-term intracell incidences. Each thick $J$-colored stub represents the nonidentity support terms of $a_J$: a bundle containing $\mathbf j\in\mathcal J_J$, with $|\mathcal J_J|=\wt(a_J)-1$. 
For each
$\mathbf j\in
\mathcal J_{\mathcal G_X(J,p)}$,
one strand connects
$X_J(u)$ to $q_p(u-\mathbf j)$,
whereas on the data side it connects
$q_p(u)$ to $X_J(u+\mathbf j)$.
For the strand bundle $\mathcal J_{{\mathcal G}_Z(J,p)}$ the signs are reversed. Arrowheads indicate the relative displacement toward the neighboring cell, not an orientation of the Tanner edge.}
\label{fig:halfcells}
\end{figure*}

Multiplying an individual $a_J$ by a group element only translates its support and gives a permutation-equivalent AMC complex.  Thus, after independent translations of the four elements, one may assume that the identity occurs in every support \cite{LinAMC2026}.  We can write
\begin{equation}
 \supp(a_J)=\{e,g_{J,1},\ldots,g_{J,r_J}\},
 \quad r_J=\wt(a_J)-1.
 \label{eq:four-supports}
\end{equation}
As in 2BGA codes, we can associate a separate, block-typed direction to every nonidentity support element,
\begin{equation}
 \cJ_J=\{\mathbf j_{J,1},\ldots,\mathbf j_{J,r_J}\},
 \qquad J\in\cD,
 \label{eq:direction-families}
\end{equation}
and define the lattice
\begin{equation}
 L
 =\bigoplus_{J\in\cD}\ \bigoplus_{\mu=1}^{r_J}\Z\mathbf j_{J,\mu}
 \cong\Z^d, \quad
 d=\sum_{J\in\cD}r_J.
 \label{eq:ambient-dimension}
\end{equation}
The support map is the group homomorphism
\begin{equation}
 \Phi: L\longrightarrow G,
 \quad
 \Phi\!\left(\sum_{J,\mu}n_{J,\mu}\mathbf j_{J,\mu}\right)
 =\prod_{J,\mu}g_{J,\mu}^{\,n_{J,\mu}}.
 \label{eq:phi-four-blocks}
\end{equation}
We define
\begin{align}
 H:=&\im\Phi\leq G, \\
 \Lambda:=&\ker\Phi.
 \label{eq:H-and-Lambda}
\end{align}
Then
\begin{equation}
 L/\Lambda\simeq H.
 \label{eq:four-block-quotient}
\end{equation}
If $H=G$, the quotient represents every group element. We then consider $u\in L/\Lambda$.  If $H<G$, the full Tanner graph decomposes into $[G:H]$ isomorphic components indexed by the cosets of $H$, and the quotient above describes one connected component.  Directions are kept block-typed even when the same group element occurs in two different supports: the resulting difference of basis vectors simply lies in $\Lambda$.  

It is useful to package the local action of a block into the two translated neighborhoods
\begin{align}
 \Nminus{J}{q_p,u}
   &:=\{q_p(u)\}\cup
      \{q_p(u-\mathbf j):\mathbf j\in\cJ_J\},
 \label{eq:nminus}\\
 \Nplus{J}{q_p,u}
   &:=\{q_p(u)\}\cup
      \{q_p(u+\mathbf j):\mathbf j\in\cJ_J\}.
 \label{eq:nplus}
\end{align}

\begin{proposition}
Let a check row and a qubit column be indexed by the same quotient lattice $L/\Lambda$.  
A nonzero block of type $J\in\mathcal D$, represented by
$\rho(a_J)$, contributes the neighborhood
$\mathcal N_J^-(q_p,u)$ to the row at $u$, whereas the transposed block
$\rho(a_J)^{\mathsf T}$ contributes
$\mathcal N_J^+(q_p,u)$.  Consequently, any block matrix whose entries are drawn from $\{0,A,B,C,D\}$ and their transposes has a geometrically local Tanner realization on $L/\Lambda$.
\end{proposition}

\emph{Reason.}  If $\rho(g)$ is the left-regular permutation matrix, then the row labeled by $h\in H$ has its nonzero column at $g^{-1}h$.  Under the identification $h=\Phi(u)$, this is the vertex $u-\mathbf j_{J,\mu}$.  Since $\rho(g)^{\mathsf T}=\rho(g^{-1})$, transposition reverses the displacement and gives $u+\mathbf j_{J,\mu}$.  The identity term stays at $u$.  Quotienting by $\ker\Phi$ identifies exactly those lattice points with the same group label.

We now consider the four-complex $Q^{(4)}$ in Eq.~\eqref{eq:complex}
and its level-2 CSS AMC4 code defined by
\begin{equation}
 H_X=Q_2^{(4)},\qquad H_Z=(Q_3^{(4)})^{\mathsf T}.
\end{equation}
The matrices can be written as~\cite{LinAMC2026}
\begin{equation}
Q_2^{(4)}=
\begin{pmatrix}
 C&B&A&0&0&0\\
 -D&0&0&B&A&0\\
 0&-D&0&-C&0&A\\
 0&0&-D&0&-C&-B
\end{pmatrix},
\label{eq:R2}
\end{equation}
\begin{equation}
(Q_3^{(4)})^{\mathsf T}=
\begin{pmatrix}
 B^{\mathsf T}&-C^{\mathsf T}&0&D^{\mathsf T}&0&0\\
 A^{\mathsf T}&0&-C^{\mathsf T}&0&D^{\mathsf T}&0\\
 0&A^{\mathsf T}&-B^{\mathsf T}&0&0&D^{\mathsf T}\\
 0&0&0&A^{\mathsf T}&-B^{\mathsf T}&C^{\mathsf T}
\end{pmatrix}.
\label{eq:R3T}
\end{equation}
Signs are irrelevant over $\F$. In addition to check-block labels in $\cD$,  
we also use the six qubit-block labels in
\begin{equation}
 \cP=\{AB,AC,AD,BC,BD,CD\}
 \label{eq:pairs}
\end{equation}
It is convenient to introduce \(\mathcal{G}_X\) and \(\mathcal{G}_Z\) as the label-valued support
tables
\begin{equation}
\begin{array}{c|cccccc}
\mathcal G_X
   & CD & BD & AD & BC & AC & AB\\
\hline
D  & C  & B  & A  & 0  & 0  & 0 \\
C  & D  & 0  & 0  & B  & A  & 0 \\
B  & 0  & D  & 0  & C  & 0  & A \\
A  & 0  & 0  & D  & 0  & C  & B
\label{eq:label-table1}
\end{array}.
\end{equation}
and
\begin{equation} 
\begin{array}{c|cccccc}
{\mathcal G}_Z & CD&BD&AD&BC&AC&AB\\ \hline
A&B&C&0&D&0&0\\
B&A&0&C&0&D&0\\
C&0&A&B&0&0&D\\
D&0&0&0&A&B&C 
\end{array}.
\label{eq:label-table2}
\end{equation}
obtained from \(Q_2^{(4)}\) and \((Q_3^{(4)})^T\), respectively, 
by replacing each nonzero block of type $J\in\mathcal D$,
namely $\rho(a_J)$ or $\rho(a_J)^{\mathsf T}$,
by its direction label $J$; the columns are naturally labeled by $p\in{\mathcal P}$.  In the explicit recursive matrices, the six qubit
columns are ordered as
\((q_{CD},q_{BD},q_{AD},q_{BC},q_{AC},q_{AB})\),
and the $X$-check rows are ordered as
\((X_D,X_C,X_B,X_A)\).
The $Z$-check rows are indexed by the
three-element subsets
\((Z_{BCD},Z_{ACD},Z_{ABD},Z_{ABC})\).
For compactness, we use the complementary single-letter
notation
$Z_A:=Z_{BCD}$,
$Z_B:=Z_{ACD}$,
$Z_C:=Z_{ABD}$,
$Z_D:=Z_{ABC}$.
Thus, the label on a $Z$-check denotes the unique direction
omitted from its associated three-element subset. This is a canonical labeling which will be useful for a general D-dimensional chain complex construction.

At every $u\in\Z^d/\Lambda$, we introduce the qubit and check fibers
\begin{align}
Q_u:&=\{q_{p}(u):p\in \mathcal P\},\\
X_u:&=\{X_J(u):J\in\mathcal D\},\\
Z_u:&=\{Z_J(u):J\in\mathcal D\}.
\end{align}
The corresponding static AMC4 unit cell is
\begin{equation}
 \mathcal E_u^{\mathrm{AMC4}}=Q_u\sqcup X_u\sqcup Z_u.
 \label{eq:AMC4-static-unit-cell}
\end{equation}
Thus every quotient vertex carries six qubits and eight checks.  Edges belonging to the identity term of a block remain within this type fiber, whereas the other support terms connect to translated copies of the appropriate qubit type.  The local Tanner incidence of the AMC4 code can be expressed as
\begin{align}
 X_J(u)
 &=\bigcup_{\substack{p\in\cP\\{\mathcal G}_X(J,p)\neq0}}
      \Nminus{{\mathcal G}_X(J,p)}{q_p,u},
 \label{eq:AMC-X-general}\\
 Z_J(u)
 &=\bigcup_{\substack{p\in\cP\\{\mathcal G}_Z(J,p)\neq0}}
      \Nplus{{\mathcal G}_Z(J,p)}{q_p,u}.
 \label{eq:AMC-Z-general}
\end{align}
For clarity, the eight check supports are
\begin{align*}
X_A(u)&=\Nminus{D}{q_{AD},u}\cup\Nminus{C}{q_{AC},u}\cup\Nminus{B}{q_{AB},u},\\
X_B(u)&=\Nminus{D}{q_{BD},u}\cup\Nminus{C}{q_{BC},u}\cup\Nminus{A}{q_{AB},u},\\
X_C(u)&=\Nminus{D}{q_{CD},u}\cup\Nminus{B}{q_{BC},u}\cup\Nminus{A}{q_{AC},u},\\
X_D(u)&=\Nminus{C}{q_{CD},u}\cup\Nminus{B}{q_{BD},u}\cup\Nminus{A}{q_{AD},u},\\
Z_A(u)&=\Nplus{D}{q_{BC},u}\cup\Nplus{C}{q_{BD},u}\cup\Nplus{B}{q_{CD},u},\\
Z_B(u)&=\Nplus{D}{q_{AC},u}\cup\Nplus{C}{q_{AD},u}\cup\Nplus{A}{q_{CD},u},\\
Z_C(u)&=\Nplus{D}{q_{AB},u}\cup\Nplus{B}{q_{AD},u}\cup\Nplus{A}{q_{BD},u},\\
Z_D(u)&=\Nplus{C}{q_{AB},u}\cup\Nplus{B}{q_{AC},u}\cup\Nplus{A}{q_{BC},u}.
\end{align*}
Figure~\ref{fig:halfcells} illustrates these incidences: thin lines show intracell connections, multistrand lines show intercell connection bundles, and the colors identify the corresponding matrix types.

The full static AMC4 code has six qubit blocks indexed by every element of $G$, and therefore
\(
 n_{\mathrm{static}}=6|G|
\).
One connected component contains $6|H|=6|L/\Lambda|$ qubits. Let $B_{\Lambda}$ be a basis matrix of $\Lambda$.  If the support elements generate $G$, then $H=G$ and the quotient-lattice model represents the full code, i.e.,
\(
N_{\rm cell}=|\det B_\Lambda|\),
\(
n_{\rm static}=6|\det B_\Lambda|
\).
The syndrome-extraction network alternates an $X$-check layer and a $Z$-check layer.  A full cell is therefore split into two types of half-cells as shown in Fig.~\ref{fig:halfcells}:
\begin{align}
 \mathfrak h_X(u)&=\{X_J(u):J\in\cD\}\ \sqcup\ \{q_p^X(u):p\in\cP\},\\
 \mathfrak h_Z(u)&=\{Z_J(u):J\in\cD\}\ \sqcup\ \{q_p^Z(u):p\in\cP\}.
\end{align}
The data nodes $q_p^X(u)$ and $q_p^Z(u)$ are consecutive ZX spiders on the same data-qubit worldline, not two independent static qubits.  In the following, we also add a temporal edge that connects the same slot $p$ between adjacent half-cells.

\subsection{Codes from a general AMC chain complex}
We now extend the preceding construction to a general \(D\)-dimensional AMC complex.
We assume $[D]:=\{1,\ldots,D\}$ and choose $D$ group-algebra elements
\begin{equation}
 a_i=e+\sum_{\mu=1}^{r_i}g_{i,\mu}\in\F[G],
 \quad A_i=\rho(a_i),
 \quad i\in[D].
 \label{eq:D-group-elements}
\end{equation}
The identity term can be included by multiplying each global AMC generator $a_i$ by a group element; this changes the complex only by permutations.  We introduce typed direction families
\begin{equation}
 \mathcal J_i=\{\mathbf j_{i,1},\ldots,\mathbf j_{i,r_i}\},
 \quad
 \Phi(\mathbf j_{i,\mu})=g_{i,\mu},
 \quad
 \Lambda=\ker\Phi,
 \label{eq:D-direction-families}
\end{equation}
where
\[
 L_D=\bigoplus_{i=1}^D\bigoplus_{\mu=1}^{r_i}\Z\mathbf j_{i,\mu},\quad d:=\sum_{i=1}^{D}|\mathcal J_i|.
\]
As before, $L_D/\Lambda\simeq H$ for the subgroup $H\leq G$ generated by the support elements.  We define qubit neighbourhoods
\begin{align}
 \mathcal N_i^-(q,u)&:=\{q(u)\}\cup\{q(u-\mathbf j):\mathbf j\in\mathcal J_i\},
 \label{eq:Di-neighborhood-minus}\\
 \mathcal N_i^+(q,u)&:=\{q(u)\}\cup\{q(u+\mathbf j):\mathbf j\in\mathcal J_i\}.
 \label{eq:Di-neighborhood-plus}
\end{align}

We consider the $D$-dimensional AMC complex
\[
 Q^{(D)}=\operatorname{AMC}(a_1,\ldots,a_D)
\]
and the level-$j$ CSS code
\begin{equation}
 H_X=Q_j^{(D)},\quad H_Z=\bigl(Q_{j+1}^{(D)}\bigr)^{\mathsf T},
 \quad 1\leq j<D.
 \label{eq:general-css}
\end{equation}
It is convenient to introduce the subset index sets
\begin{equation}
 \mathsf S_k:=\{I\subseteq[D]:|I|=k\}.
 \label{eq:subset-index-sets}
\end{equation}
In the canonical ordering, row blocks of $Q_j^{(D)}$ are indexed by $K\in\mathsf S_{j-1}$, column blocks by $I\in\mathsf S_j$, and column blocks of $Q_{j+1}^{(D)}$ by $L\in\mathsf S_{j+1}$.  Up to the orientation signs, which are irrelevant over $\F$, the block entries are
\begin{align}
 (Q_j^{(D)})_{K,I}
 &=\begin{cases}
   \pm A_i,& I=K\cup\{i\},\\
   0,&\text{otherwise},
   \end{cases}
 \label{eq:Qj-subset-blocks}\\
 (Q_{j+1}^{(D)})_{I,L}
 &=\begin{cases}
   \pm A_i,& L=I\cup\{i\},\\
   0,&\text{otherwise}.
   \end{cases}
 \label{eq:Qjp1-subset-blocks}
\end{align}

\begin{definition}[Static level-$j$ unit cell]
For every $u\in L_D/\Lambda$, we define the unit cell as
\begin{align}
 \mathcal E_u^{(D,j)}
 =&\{q_I(u):I\in\mathsf S_j\} \nonumber \\ 
 &\sqcup\{X_K(u):K\in\mathsf S_{j-1}\}\nonumber \\ 
 &\sqcup\{Z_L(u):L\in\mathsf S_{j+1}\}.
 \label{eq:general-chain-unit-cell}
\end{align}
Thus the unit cell contains $\binom Dj$ qubit types,
 $\binom D{j-1}$ $X$-check types, and
 $\binom D{j+1}$ $Z$-check types.
 \label{eq:general-unit-cell-counts}
\end{definition}

The exact local incidence rules are
\begin{align}
 X_K(u):&
 \bigcup_{i\in[D]\setminus K}
 \mathcal N_i^-\!\left(q_{K\cup\{i\}},u\right),
 \quad K\in\mathsf S_{j-1},
 \label{eq:general-chain-X-incidence}\\
 Z_L(u):&
 \bigcup_{i\in L}
 \mathcal N_i^+\!\left(q_{L\setminus\{i\}},u\right),
 \qquad L\in\mathsf S_{j+1}.
 \label{eq:general-chain-Z-incidence}
\end{align}
In an $X$ row, the block $A_i$ connects the check type $K$ to the qubit type $K\cup\{i\}$ in the cell itself and in all negative $i$-directions.  In a $Z$ row, the transposed block $A_i^{\mathsf T}$ connects the check type $L$ to the qubit type $L\setminus\{i\}$ in the cell itself and in all positive $i$-directions.
We use the canonical subset labeling of the AMC4 block
spaces. 

\begin{proposition}
The Tanner graph obtained by translating the unit cell~\eqref{eq:general-chain-unit-cell} over $L_D/\Lambda$ and adding the edges in \eqref{eq:general-chain-X-incidence}--\eqref{eq:general-chain-Z-incidence} is permutation-isomorphic to the Tanner graph of AMC code in \eqref{eq:general-css}.  If $H<G$, the full code is the disjoint union of identical coset-indexed copies of this lattice graph.
\end{proposition}

\emph{Reason.}  The subset relations in \eqref{eq:Qj-subset-blocks} and \eqref{eq:Qjp1-subset-blocks} identify exactly which check and qubit types are joined by $A_i$ or $A_i^{\mathsf T}$.  The regular-representation dictionary sends $A_i$ to the negative-direction neighborhood $\mathcal N_i^-$ and $A_i^{\mathsf T}$ to the positive-direction neighborhood $\mathcal N_i^+$.  Quotienting by $\Lambda$ identifies precisely equal group labels.

Let $B_{\Lambda}$ be a basis matrix of $\Lambda$. One connected component contains $\binom Dj |H|= \binom Dj |L_D/\Lambda|$ qubits.  If the support elements generate $G$, then $H=G$ and the quotient-lattice model represents the full code, i.e.,
\begin{equation}
N_{\rm cell}=|\det B_{\Lambda}|,
\qquad
n_{\rm static}=\binom Dj|\det B_{\Lambda}|.
\end{equation}

The syndrome-extraction unit cell is split into two typed half-cells
\begin{align}
 \mathfrak h_X^{(D,j)}(u)
 &=\{X_K(u):K\in\mathsf S_{j-1}\}
   \sqcup\{q_I^X(u):I\in\mathsf S_j\},
 \label{eq:general-X-halfcell}\\
 \mathfrak h_Z^{(D,j)}(u)
 &=\{Z_L(u):L\in\mathsf S_{j+1}\}
   \sqcup\{q_I^Z(u):I\in\mathsf S_j\}.
 \label{eq:general-Z-halfcell}
\end{align}
The two copies $q_I^X(u)$ and $q_I^Z(u)$ are consecutive data spiders on the same worldline and are connected by temporal legs. Therefore, the $X$ half-cell contains $\binom D{j-1}$ check spiders and $\binom Dj$ data spiders, while the $Z$ half-cell contains $\binom D{j+1}$ check spiders and $\binom Dj$ data spiders.
As a result, we can state that the number of checks in a half-cell equals the number of row blocks of the corresponding check matrix, while the number of data slots equals its number of column blocks.

In the cyclic case, one can explicitly determine the number of encoded qubits~\cite{LinAMC2026}. We define the characteristic polynomial
\[
h(x)
=
\gcd\!\left(
a_1(x),\ldots,a_D(x),x^\ell-1
\right),
\,
\kappa:=\deg h(x).
\]
Since
\[
\operatorname{rank}Q_r^{(D)}
=
\binom{D-1}{r-1}(\ell-\kappa),
\]
the number of encoded qubits at level $j$ reduces to
\begin{equation}
k_j=\binom{D}{j}\kappa.
\label{eq:cyclic-amc-encoded-qubits}
\end{equation}

\section{Lift to spacetime and Floquet construction}
We consider a general \(D\)-dimensional AMC complex and its level-$j$ AMC code.
To apply the direct Floquetification procedure of Ref.~\cite{Stairway2026} to the
foliated CSS ZX network, every spider must have even valence. Thus, every row and every column of both
$H_X$ and $H_Z$ must have even Hamming weight. 
Define $\omega_i:=\wt(a_i)$.  The spider degrees follow directly from the subset incidence:
\begin{align}
 \deg X_K&=\sum_{i\notin K}\omega_i,
 &
 \deg q_I^X&=2+\sum_{i\in I}\omega_i,
 \label{eq:general-X-degrees-nonuniform}\\
 \deg Z_L&=\sum_{i\in L}\omega_i,
 &
 \deg q_I^Z&=2+\sum_{i\notin I}\omega_i.
 \label{eq:general-Z-degrees-nonuniform}
\end{align}
The additional two legs are the temporal input and output legs.  A direct worldline matching requires every spider to have even degree and, after a time covector is chosen, equal numbers of incoming and outgoing ports.  If all block weights are equal to $\omega$, these formulas reduce to
\begin{align}
 \deg(X_K)&=(D+1-j)\omega, \quad
 \deg q_I^X=j\omega+2,\\
 \deg(Z_L)&=(j+1)\omega, \quad
 \deg q_I^Z=(D-j)\omega+2.
 \label{eq:general-degrees}
\end{align}
For $D=4$ and $j=2$, the subset-indexed cell has six qubit types, four $X$-check types, and four $Z$-check types.  For weight-two elements, all four degrees in \eqref{eq:general-X-degrees-nonuniform} and \eqref{eq:general-Z-degrees-nonuniform} equal six, so every spider is a $3\to3$ map. 

We introduce the original syndrome-extraction time direction $\mathbf j_0$ and work on
\begin{equation}
\widetilde L_{d+1}
:=
\operatorname{span}_{\mathbb Z}\!\left(
\{\mathbf j_0\}
\sqcup
\bigsqcup_{i=1}^{D}\mathcal J_i
\right)
\cong
\mathbb Z^{d+1}.
\end{equation}
To implement syndrome extraction as a Floquet schedule, we choose the rotated-time covector
\begin{equation}
 \mathbf t=(2,1,\ldots,1).
 \label{eq:time-covector}
\end{equation}
To avoid placing two half-cells at the same lattice site, the timing convention depends on whether the half-cell is $Z$- or $X$-type:
\begin{equation}
 T_Z(u')=\mathbf t\cdot u',\qquad
 T_X(u')=\mathbf t\cdot u'+1,
 \label{eq:halfcell-times}
\end{equation}
where $u'\in\Z^{d+1}$.
Thus a step along any abstract spatial basis vector advances one unit of rotated time, while a full step along $\mathbf j_0$ advances two units, i.e., one unit per half-cell.

Qubits in the rotated-time \(T=0\) layer evolve according to Floquet boundary conditions defined by the matrix $B_{\Lambda'}$, and the spacetime code is the quotient $\Z^{d+1}/\Lamp$. Note that fixed-rotated-time cuts of this quotient are finite. The spacetime quotient must not identify distinct times across the boundary.  Therefore every periodicity vector $v'\in\Lambda'$ must satisfy
\begin{equation}
 v'\cdot\mathbf t=0.
 \label{eq:no-ctc}
\end{equation}
A canonical doubled lift of a spatial vector of the static code, $v=(v_1,\ldots,v_d)\in\Lambda$, is
\begin{equation}
 {\mathcal F}(v)=\bigl(-\textstyle\sum_{i=1}^d v_i,\ 2v_1,\ldots,2v_d\bigr).
 \label{eq:doubled-lift}
\end{equation}
Indeed, $\mathbf t\cdot {\mathcal F}(v)=0$ identically.

A Floquet worldline assignment can be constructed using the following steps:
\begin{enumerate}
 \item Assign $X$ spiders to the $X$ checks and $Z$ spiders to the $q_I^X$ data nodes within the $X$ half-cell. Use the opposite assignment in the $Z$ half-cell.
 \item Orient each incident leg as incoming or outgoing according to the sign of its rotated-time displacement, and add $\tau_{\rm in}$ and $\tau_{\rm out}$ ports to each data qubit.
 \item Require equal numbers of incoming and outgoing legs at every spider.
 \item Choose a bijection (port matching) from the incoming legs to the outgoing legs of each spider.
 \item Trace the matched ports through adjacent half-cells to obtain physical-qubit worldlines.
 \item Decompose each even-legged spider into two-qubit $XX$ or $ZZ$ parity measurements along the chosen matching.
\end{enumerate}
We note that different matchings produce different worldline cycles and potentially different finite Floquet codes, even when the static matrices and periodicity lattice are fixed.

The number of half-cells in the $T=0$ layer determines the number of worldlines, or physical qubits, $n_{\rm floq}$, in a Floquet code. To calculate the number of half-cells in this layer, let $B_{\Lambda'}\in\operatorname{Mat}_{d\times(d+1)}
(\mathbb Z)$ have rows forming a basis of $\Lambda'$.
Using Eq.~\eqref{eq:time-covector}, define
\begin{equation}
\widehat B_{\Lambda',\mathbf t}
:=
\begin{pmatrix}
B_{\Lambda'}\\
\mathbf t
\end{pmatrix},
\end {equation}
which leads to
\begin{equation}
N_{\rm cell}'
=
\frac{|\det \widehat B_{\Lambda',\mathbf t}|}{\|\mathbf t\|_2^2}.
\end{equation}
Let $n_{\rm in}$ denote the number of incoming
worldline ports per half-cell. Then the number of qubits in the Floquet code is
\begin{equation}
n_{\rm floq}=n_{\rm in}\frac{|\det \widehat B_{\Lambda',\mathbf t}|}{\|\mathbf t\|_2^2}.
\label{eq:nfloq}
\end{equation}

\section{Examples of Floquet constructions from AMC codes}
\subsection{GB Floquet construction}
Here we consider GB codes constructed from weight-two circulant matrices~\cite{WangPryadko2022}. Formally, this case corresponds to the two-dimensional AMC complex. The static GB CSS code is
defined by
\begin{equation}
A=I+P_1,\qquad B=I+P_m.
\label{eq:A-B}
\end{equation}
The static code has two data-qubit columns per group element. We write them as $q_A(r)$ and $q_B(r)$, $r\in\Z_\ell$. With the above convention, the exact matrix incidences are
\begin{align}
X(r)&:\quad q_A(r),\;q_A(r-1),\;q_B(r),\;q_B(r-m),
\label{eq:X-cyclic-incidence}\\
Z(r)&:\quad q_A(r),\;q_A(r+m),\;q_B(r),\;q_B(r+1).
\label{eq:Z-cyclic-incidence}
\end{align}

Let $u_A$ and $u_B$ denote the two spatial displacements. We then have
\begin{equation}
\pi:\Z^2\longrightarrow \Z_\ell,
\quad
\pi(u_A,u_B)=u_A+m u_B\pmod \ell.
\label{eq:phi}
\end{equation}
Translation by the abstract $j_A$ direction changes the cyclic index by 1, and translation by the abstract $j_B$ direction changes the cyclic index by $m$. The finite cyclic code is the quotient
$\Z^2/\Lam$,
$\Lam=\ker \pi$.
A convenient row basis is
\begin{equation}
\Lam=\left\langle (\ell,0),\;(-m,1)\right\rangle.
\label{eq:spatial-lambda}
\end{equation}

\begin{figure*}[t]
\centering
\includegraphics{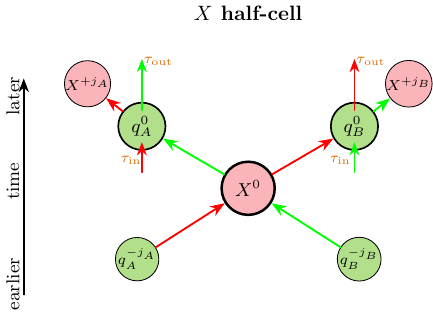}%
\hfill
\includegraphics{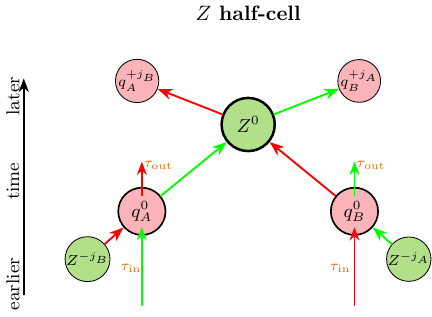}
\caption{Local incidence diagram for a GB code with weight-four checks. A unit cell containing two qubits and two checks is split into two half-unit cells of $X$- and $Z$-type. Each half-unit cell has $n_{\rm in}=4$ inputs. Within each half-unit cell, each data node is turned into a spider by adding extra $\tau_{\rm in}$ and $\tau_{\rm out}$ ports. Local objects are labeled with superscript $(0)$, while external ports are labeled by relative lattice directions such as $q^{-j_A}$ and $X^{+j_B}$. The colors mark the chosen port matching for the corresponding check--data ZX network. In addition, $X$ spiders are marked by pink and $Z$ spiders are marked by green.}
\label{fig:matrix-exact-incidence}
\end{figure*}

In the $\Z^2/\Lam$ notation the exact local incidence relations become
\begin{align}
X(u)&:\quad q_A(u),\;q_A(u-\mathbf j_A),\;q_B(u),\;q_B(u-\mathbf j_B),
\label{eq:X-Z2-incidence}\\
Z(u)&:\quad q_A(u),\;q_A(u+\mathbf j_B),\;q_B(u),\;q_B(u+\mathbf j_A).
\label{eq:Z-Z2-incidence}
\end{align}
Thus, we define a unit cell containing two checks and two qubits on the $\Z^2/\Lam$ lattice~\cite{Arnault2026}; see Fig.~\ref{fig:matrix-exact-incidence}.

Following the Floquetification procedure~\cite{Stairway2026}, we introduce a spacetime lattice with one original syndrome-extraction time coordinate and two abstract spatial coordinates:
\[
u^\prime=(u_0,u_A,u_B)\in\Z^3.
\]
To implement syndrome extraction as a Floquet schedule, we choose the rotated-time covector~\cite{Stairway2026}
\begin{equation}
t=(2,1,1),
\qquad
T(u^\prime)=t\cdot u^\prime=2u_0+u_A+u_B.
\label{eq:rot-time}
\end{equation}
We use the standard timing convention for $Z$- and $X$-type half-cells, $T_Z(u^\prime)=T(u^\prime)$,
and $T_X(u^\prime)=T(u^\prime)+1$.
Thus, a rotated-time \(T=0\) layer contains the $Z$ half-cells with base time $T=0$ and the $X$ half-cells with base time $T=-1$.

Periodic boundary vectors must be orthogonal to $t$ so that the quotient has no closed timelike curves~\cite{Stairway2026}. A canonical doubled lift is
\begin{equation}
{\mathcal F}:\Z^2\longrightarrow \Z^3,
\quad
{\mathcal F}(a,b)=(-a-b,2a,2b).
\label{eq:doubled-lift-2d}
\end{equation}
It obeys $t\cdot {\mathcal F}(a,b)=0$.
Applying \eqref{eq:doubled-lift-2d} to the basis \eqref{eq:spatial-lambda} gives
\begin{equation}
B_{\Lamp}
=\begin{bmatrix}
-\ell & 2\ell & 0\\
m-1 & -2m & 2
\end{bmatrix}.
\label{eq:lambda-prime-general}
\end{equation}

We now construct the spacetime code as the quotient $\Z^3/\Lamp$, represented by the lattice of half-unit cells shown in Fig.~\ref{fig:matrix-exact-incidence}. Its local connectivity is determined by the static code.
Within each half-unit cell, each qubit node is turned into a spider by adding extra $\tau_{\rm in}$ and $\tau_{\rm out}$ ports as shown in Fig.~\ref{fig:matrix-exact-incidence}.
We orient each incident leg as incoming or outgoing according to the sign of its rotated-time displacement. 
To trace the qubit worldlines, we choose a port matching, which also determines the pairwise-measurement schedule after applying the rules in Fig.~\ref{fig:stairway-style-decomposition}. We use the port matching shown in Fig.~\ref{fig:matrix-exact-incidence}. The measurement schedule begins with a layer of half-unit cells at $T=0$ and then develops along the rotated-time direction, with the physical qubits represented by the worldlines crossing the rotated-time cut, modulo periodic boundaries determined by $\Lamp$. The number of worldlines is determined by Eq.~\eqref{eq:nfloq}, with $n_{\rm in}=4$ inputs per half-cell for the GB codes considered here.

Using Fig.~\ref{fig:matrix-exact-incidence}, we can trace the green worldline 
\begin{equation}
\tau_{\rm in} \to j_A \to \tau_{\rm out} \to \tau_{\rm in} \to j_B \to \tau_{\rm out}, 
\end{equation}
and the red worldline
\begin{equation}
\tau_{\rm in} \to j_B \to \tau_{\rm out} \to \tau_{\rm in} \to j_A \to \tau_{\rm out}, 
\end{equation}
from which we see that the local period is $T_{\rm local}=\|t\|_1=4$. Over the course of such a local round, each qubit undergoes a constant displacement $\vec{\delta}$ in the three-dimensional spacetime lattice, after which the local circuit repeats. This local period differs from the global period $T_{\rm glob}$, after which each qubit returns to its initial position; the latter is largely determined by $\Lambda^\prime$. Table~\ref{tab:2bga-weight4-floquet} lists the parameters of Floquet codes constructed from GB codes for several $(\ell,m)$ pairs.

\begin{table}[t]
\centering
\small
\begin{tabular}{c c c c r r r r r r c}
\toprule
$\ell$  & $m$ & $N_{\rm cell}'$ & $n_{\rm floq}$ &
$k_{\rm floq}$ & $\mathrm{rank}H_X^{\rm ISG}$&$\mathrm{rank}H_Z^{\rm ISG}$ & $(d_X,d_Z)$\\
\midrule
7 &  $3$ & $14$ & 56 &  2 & $20$&$34$ & $(5,5)$\\
11 &  $3$ & $22$ & 88  & 2 & $32$&$54$ & $(5,5)$\\
13 &  $5$ & $26$ & 104  & 2 & $38$&$64$ & $(8,8)$\\
17 &  $4$ & $34$ & 136  & 2 & $50$&$84$ & $(9,9)$\\
\bottomrule
\end{tabular}
\caption{Floquet versions of weight-four GB codes containing $n_{\rm floq}$ physical qubits and encoding $k_{\rm floq}$ obtained using the rotated-time construction. The displayed ranks and distances are evaluated at
integer-layer cuts and are independent of the chosen
integer cut for the examples listed. At fractional times, the ranks and distances may change and, at certain cuts, interchange between the $X$ and $Z$ sectors, with $\operatorname{rank}({\rm ISG})=\operatorname{rank}H_X^{\rm ISG}+\operatorname{rank}H_Z^{\rm ISG}$ staying constant.}
\label{tab:2bga-weight4-floquet}
\end{table}

\begin{figure*}[t]
\centering
\includegraphics{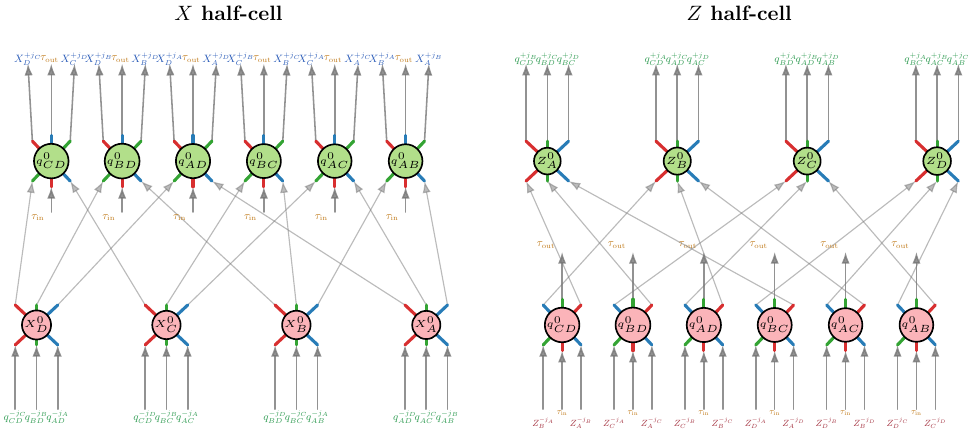}
\caption{Local incidence diagram for an AMC code with weight-six checks. A unit cell containing six qubits and eight checks is split into two half-unit cells of $X$- and $Z$-type. Each half-unit cell has $n_{\rm in}=18$ inputs. Within each half-unit cell, each data node is turned into a spider by adding extra $\tau_{\rm in}$ and $\tau_{\rm out}$ ports. Local objects are labeled with superscript $(0)$, while external ports are labeled by relative lattice directions such as $q^{-j_A}_{AB}$ and $X^{+j_D}_C$. The colors mark the chosen port matching for the corresponding check--data ZX network and gray lines show the full local connections. In addition, $X$ spiders are marked by pink and $Z$ spiders are marked by green.}
\label{fig:amc-matching-worldlines}
\end{figure*}

\subsection{AMC4 Floquet construction}
Here, we consider $\mathcal Q^{(4)}
=\operatorname{AMC}(a_A,a_B,a_C,a_D)$ with
\begin{equation}
a_A, a_B, a_C, a_D \in \Ftwo[C_\ell].
\end{equation}
If each group-algebra element has weight two,
\begin{equation}
 a_J=e+g_J,
 \quad \rho(a_J)=I+\rho(g_J),
 \quad J\in\cD,
 \label{eq:weight-two-group-elements}
\end{equation}
then each family has one vector and $d=4$:
\[
 \cJ_A=\{\mathbf j_A\},\,
 \cJ_B=\{\mathbf j_B\},\,
 \cJ_C=\{\mathbf j_C\},\,
 \cJ_D=\{\mathbf j_D\}.
\]
We identify the cyclic group
$C_\ell=\langle x\mid x^\ell=e\rangle$
with the additive group $\mathbb Z_\ell$ through
$x^r\leftrightarrow r$.
This leads to a construction based on weight-two circulant matrices
\begin{equation}
A=I+P_{s_A}, B=I+P_{s_B}, C=I+P_{s_C}, D=I+P_{s_D},
\end{equation}
where each cell is labeled by
\[
u=(u_A,u_B,u_C,u_D)\in\mathbb Z^4.
\]

The four abstract translations are projected to a single cyclic coordinate
$r\in\mathbb Z_\ell$
by a homomorphism $\pi:\mathbb Z^4\longrightarrow \mathbb Z_\ell$.
\begin{equation}
r=\pi(u)
=
s_A u_A+s_B u_B+s_C u_C+s_D u_D
\pmod{\ell},
\label{eq:amc-phi-map}
\end{equation}
where $r$ is a cyclic coordinate for both checks and qubits. 
Translating by $\mathbf j_A$, $\mathbf j_B$, $\mathbf j_C$, $\mathbf j_D$ changes the cyclic matrix index by
$s_A$, $s_B$, $s_C$, $s_D$, respectively.
The kernel of this map defines the periodicity lattice
\begin{align}
\Lambda=&\bigl\{
\mathbf v\in\mathbb Z^4:\\ \nonumber
& s_Av_A+s_Bv_B+s_Cv_C+s_Dv_D=0\pmod{\ell}
\bigr\}.
\label{eq:amc-lambda-kernel}
\end{align}
When the shifts generate all of \(\mathbb Z_\ell\), the quotient satisfies
$\mathbb Z^4/\Lambda\simeq \mathbb Z_\ell$. The resulting half-cells and the Tanner graph corresponding to Eqs.~\eqref{eq:AMC4-static-unit-cell}, \eqref{eq:AMC-X-general}, and \eqref{eq:AMC-Z-general} are shown in Fig.~\ref{fig:amc-matching-worldlines}.
\begin{table*}[t]
\centering
\small
\renewcommand{\arraystretch}{1.25}
\begin{tabular}{c c c}
\toprule
\(q_p\)
&
\(\tau_X(q_p)\)-trajectory
&
\(\tau_Z(q_p)\)-trajectory
\\
\midrule
\(q_{CD}\)
&
\(+j_C\to +j_D\to \tau\to +j_B\to +j_A\to \tau\)
&
\(+j_B\to +j_A\to \tau\to +j_C\to +j_D\to \tau\)
\\
\(q_{BD}\)
&
\(+j_B\to +j_D\to \tau\to +j_C\to +j_A\to \tau\)
&
\(+j_C\to +j_A\to \tau\to +j_B\to +j_D\to \tau\)
\\
\(q_{AD}\)
&
\(+j_A\to +j_D\to \tau\to +j_C\to +j_B\to \tau\)
&
\(+j_C\to +j_B\to \tau\to +j_A\to +j_D\to \tau\)
\\
\(q_{BC}\)
&
\(+j_B\to +j_C\to \tau\to +j_D\to +j_A\to \tau\)
&
\(+j_D\to +j_A\to \tau\to +j_B\to +j_C\to \tau\)
\\
\(q_{AC}\)
&
\(+j_A\to +j_C\to \tau\to +j_D\to +j_B\to \tau\)
&
\(+j_D\to +j_B\to \tau\to +j_A\to +j_C\to \tau\)
\\
\(q_{AB}\)
&
\(+j_A\to +j_B\to \tau\to +j_D\to +j_C\to \tau\)
&
\(+j_D\to +j_C\to \tau\to +j_A\to +j_B\to \tau\)
\\
\bottomrule
\end{tabular}
\caption{
Contracted local worldline trajectories for the AMC4 matching.
Each row gives the two possible starts, \(\tau_X(q_p)\) and \(\tau_Z(q_p)\).
After six steps, the worldline returns to the same local port type,
possibly translated in the absolute spacetime lattice. The symbol $\tau$ denotes the contracted
temporal segment
$\tau_{\mathrm{out}}\to\tau_{\mathrm{in}}$
between consecutive half-cells.
}
\label{tab:amc4-worldline-trajectories}
\end{table*}

When \(s_A=1\), a convenient row basis is
\[
B_{\Lambda}=
\begin{pmatrix}
\ell & 0 & 0 & 0\\
-s_B & 1 & 0 & 0\\
-s_C & 0 & 1 & 0\\
-s_D & 0 & 0 & 1
\end{pmatrix}.
\]
It follows from Eq.~\eqref{eq:cyclic-amc-encoded-qubits} that
for the level-2 code the number of encoded qubits is
\[
k
=
n_{\rm static}
-\operatorname{rank}H_X
-\operatorname{rank}H_Z
=
6\kappa.
\]

The spacetime lattice is
\[
\widetilde L_5
=
\langle \mathbf j_0,\mathbf j_A,\mathbf j_B,\mathbf j_C,\mathbf j_D\rangle,
\]
and the rotated-time covector is
\(
\mathbf t=(2,1,1,1,1)
\).
Here $u^\prime\in\mathbb Z^5$, and the half-cell times are defined by Eq.~\eqref{eq:halfcell-times}.

Qubits in the rotated-time \(T=0\) layer evolve according to Floquet boundary conditions defined by a rank-four lattice $\Lambda'\subset \mathbb Z^5$.
To avoid closed timelike
curves and make \(T\) well-defined on the quotient, we require \eqref{eq:no-ctc}.
A convenient choice is the canonical doubled lift
\[
{\mathcal F}(v)=
\bigl(
-v_A-v_B-v_C-v_D,\,
2v_A,2v_B,2v_C,2v_D
\bigr).
\]
The rows of the matrix $B_{\Lambda}$ may be used in this lift.

We construct the spacetime code as the quotient $\Z^5/\Lamp$, represented by the lattice of half-unit cells shown in Fig.~\ref{fig:amc-matching-worldlines}.
Within each half-unit cell, each qubit node is turned into a spider by adding extra $\tau_{\rm in}$ and $\tau_{\rm out}$ ports.
The qubit lines follow the port matching shown in Fig.~\ref{fig:amc-matching-worldlines}. The measurement schedule begins with a layer of half-unit cells at $T=0$ and then develops along the rotated-time direction, with the physical qubits represented by the worldlines crossing the rotated-time cut, modulo periodic boundaries determined by $\Lamp$. The number of worldlines is determined by Eq.~\eqref{eq:nfloq}, with $n_{\rm in}=18$ inputs per half-cell for the AMC4 code considered here.

Taking $\ell=10$ and $(s_A,s_B,s_C,s_D)=(1,2,3,4)$,
we can use
\[
B_{\Lambda'}=
\begin{pmatrix}
1&-4&2&0&0\\
-5&10&0&0&0\\
1&-3&0&1&0\\
2&-6&1&0&1
\end{pmatrix},
\]
which results in $N_{\rm cell}'=10$ and $n_{\rm floq}=180$; see Table~\ref{tab:floquet-amc4-tableI} for additional code examples.

For a construction based on weight-two circulant matrices, all spiders have six legs as shown in Fig.~\ref{fig:amc-matching-worldlines}.  Each local spider requires a port matching
\begin{equation}
  \mu:\{\text{incoming legs at }{u^\prime}\}\to \{\text{outgoing legs at }{u^\prime}\}.
\end{equation}
The port matching determines the qubit worldlines, and the pairwise-measurement circuit in Fig.~\ref{fig:stairway-style-decomposition} is applied only after choosing the port matching of each high-degree spider.
For $\sigma\in\{X,Z\}$, we define the incidence sets
\begin{align}
\mathcal P_\sigma(J)
&:=
\left\{
p\in\mathcal P:
\mathcal G_\sigma(J,p)\neq0
\right\},
\qquad J\in\mathcal D,
\\
\mathcal D_\sigma(p)
&:=
\left\{
J\in\mathcal D:
\mathcal G_\sigma(J,p)\neq0
\right\},
\qquad p\in\mathcal P.
\end{align}
The check-spider ports are indexed by
$p\in\mathcal P_\sigma(J)$, whereas the spatial ports of a
data spider are indexed by
$J\in\mathcal D_\sigma(p)$; the temporal port is denoted by
$\tau$. The same index is used for the corresponding
incoming and outgoing ports. Hence, a local matching is a
permutation of the port-label set itself:
\begin{align}
\mu_\sigma^{\mathrm{check}}(J)
&\in
\operatorname{Sym}\!\left(\mathcal P_\sigma(J)\right),
\\
\mu_\sigma^{\mathrm{data}}(p)
&\in
\operatorname{Sym}\!\left(
\{\tau\}\sqcup\mathcal D_\sigma(p)
\right).
\end{align}
An incoming port labeled by $\alpha$ is matched to the
outgoing port labeled by $\mu(\alpha)$.

For the matching shown in Fig.~\ref{fig:amc-matching-worldlines},
the check-spider matchings are trivial:
\begin{equation}
\mu_X^{\mathrm{check}}(J)
=
\operatorname{id}_{\mathcal P_X(J)},
\qquad
\mu_Z^{\mathrm{check}}(J)
=
\operatorname{id}_{\mathcal P_Z(J)}.
\end{equation}
For a qubit label $p=I_1I_2\in\mathcal P$, where
$I_1,I_2\in\mathcal D$ and $I_1<I_2$ in the order $A<B<C<D$, let
$\{K_1,K_2\}=\mathcal D\setminus\{I_1,I_2\}$ with 
$K_1<K_2$.
The data-spider matchings are
\begin{equation}
\mu_X^{\mathrm{data}}(p)
=
(\tau\,I_2\,I_1),
\qquad
\mu_Z^{\mathrm{data}}(p)
=
(\tau\,K_1\,K_2),
\label{eq:amc4-port-matchings}
\end{equation}
where $(a\,b\,c)$ denotes the cycle
$a\mapsto b\mapsto c\mapsto a$.
Using these matching rules, Table~\ref{tab:amc4-worldline-trajectories} tracks trajectories that start at the $\tau_{\rm in}$ port of a qubit labeled by $p$. For the chosen matching, we identify the local period $T_{\rm local}=6$ after which the worldline returns to the same local input position within the unit cell, while its absolute coordinate is displaced by $\vec{\delta}$ in the five-dimensional spacetime lattice.
We can also identify the global period $T_{\rm glob}$, after which each qubit returns to its initial global position; see Table~\ref{tab:floquet-amc4-tableI}. This period is largely determined by $\Lambda^\prime$.

\begin{table*}[t]
\centering
\small
\setlength{\tabcolsep}{3.5pt}
\begin{tabular}{c c r r c r r r r r r r c c}
\toprule
$(\ell,s_A,s_B,s_C,s_D)$ & $n_{\rm static}$ & $d_{\rm static}$ &
$N_{\rm cell}'$ & $n_{\rm floq}$ & $T_{\rm glob}$  &
$k_{\rm floq}$ &
$\operatorname{rank}H_X^{\rm ISG}$ & $\operatorname{rank}H_Z^{\rm ISG}$ &
$(d_X,d_Z)$ & $d_{\rm emb}$ & $kd^2_{\rm emb}/n$  \\
\midrule
$(7,1,2,3,4)$    & 42  & 4  & $14$   & 252 & 42  & 6 & 67  & 179 & $(9,9)$  & $9$ & $1.93$\\ 
$(10,1,2,3,4)$    & 60  & 5  & $10$   & 180 & 6   & 6 & 47  & 127 & $(8,8)$  & $8$ & $2.13$\\
$(11,1,2,3,4)$    & 66  & 6  & $22$ & 396 & 66  & 6 & 107 & 283 & $(10,10)$ & $10$ & $1.52$ \\
$(12,1,3,5,7)$    & 72  & 5  & $6$ & 108 & 18  & 6 & 27 & 75 & $(5,5)$ & $5$ & $1.34$ \\
$(14,1,2,5,6)$    & 84  & 7  & $14$   & 252 & 6   & 6 & 67  & 179 & $(8,8)$ & $8$ & $1.52$ \\
$(16,1,3,5,7)$    & 96  & 8  & $8$   & 144 & 6   & 6 & 37  & 101 & $(8,8)$ & $8$ & $2.67$ \\
$(18,1,3,5,7)$    & 108 & 9  & $18$   & 324 & 54   & 6 & 87  & 231 & $(10,10)$ & $10$ & $1.85$\\
$(25,1,4,6,9)$    & 150 & 10 & $50$ & 900 & 30   & 6 & 247 & 647 & $(14,\le14)$ & $\le14$ & $1.31$\\
$(28,1,3,7,12)$   & 168 & 11 & $28$ & 504 & 84   & 6 & 137 & 361 & $(12,10)$  & $10$ & $1.19$\\
$(30,2,5,8,9)$    & 180 & 12 & $30$ & 540 & 30   & 6 & 147 & 387 & $(15, 15)$ & $15$ & $2.5$\\
\bottomrule
\end{tabular}
\caption{Floquet AMC4 codes obtained from the $T=0$ layer of half-cells. The layer is evolved modulo $\Lambda'$. Rather than using the doubled lift ${\mathcal F}(v)$, we choose $\Lambda'$ to minimize $n_{\rm floq}$. The distances are calculated using MILP\@. The distances $d_X$ and $d_Z$ and the matrix ranks are evaluated at integer values of $T$. The CSS ranks and ISG distances are independent of the integer cut $T$ for
the examples listed.  At fractional times, the ranks and distances may change and, at certain cuts, interchange between the $X$ and $Z$ sectors, with $\operatorname{rank}({\rm ISG})=\operatorname{rank}H_X^{\rm ISG}+\operatorname{rank}H_Z^{\rm ISG}$ staying constant. }
\label{tab:floquet-amc4-tableI}
\end{table*}
A particular realization of a Floquet code, including the number of encoded qubits, depends on the choice of matching. A valid matching must ensure that each retained geometric
edge has positive time orientation under $\mathbf t$ and
that the resulting worldline displacement is compatible
with the periodic identifications generated by $\Lambda'$. Table~\ref{tab:floquet-amc4-tableI} lists the parameters of Floquet codes constructed from the AMC4 codes of Ref.~\cite{LinAMC2026}.

\section{Numerical results}
\subsection{Distance calculation by mixed-integer linear programming}
The code distance was calculated using mixed-integer
linear programming (MILP), following the standard formulation
for stabilizer codes and its extension to dynamical and
subsystem codes~\cite{Landahl2011ColorCodes,Stairway2026}.
For each Pauli sector $P\in\{X,Z\}$ and time slice $t$, let
\[
\bigl\{\boldsymbol{\ell}^{(P,t)}_q\bigr\}_{q=1}^{k}
\]
be a basis of protected, non-gauge logical operators, and let
\[
\bigl\{\mathbf{s}^{(P,t)}_r\bigr\}_{r=1}^{m_P}
\]
be a generating set for the corresponding stabilizer subgroup.
Every representative of a nontrivial logical class can then be
written in binary form as
\[
\mathbf{v}
=
\bigoplus_{q=1}^{k}
\beta_q\boldsymbol{\ell}^{(P,t)}_q
\oplus
\bigoplus_{r=1}^{m_P}
\alpha_r\mathbf{s}^{(P,t)}_r,
\]
where $\alpha_r,\beta_q\in\{0,1\}$ and at least one
$\beta_q$ is nonzero.

The minimum-weight representative is obtained from
\begin{align}
d_P(t)=\min\quad&
\sum_{i=1}^{n}x_i,
\label{eq:distance-milp-objective}\\
\text{subject to}\quad&
\sum_{q=1}^{k}\beta_q\boldsymbol\ell^{(P,t)}_{q}
+
\sum_{r=1}^{m_P}\alpha_r \mathbf{s}^{(P,t)}_{r}
-2\boldsymbol \eta
=\mathbf{x},
\label{eq:distance-milp-parity}\\
&
\sum_{q=1}^{k}\beta_q\geq 1,
\label{eq:distance-milp-logical}\\
&
x_i,\alpha_r,\beta_q\in\{0,1\},
\qquad
\eta_i\in\mathbb Z_{\geq0}.
\end{align}
The integer vector $\boldsymbol \eta$ implements addition modulo two,
while the vector $\mathbf{x}$ records whether the resulting logical operator
acts nontrivially on a qubit.  For a static CSS code,
\[
d_{\rm static}=\min\{d_X,d_Z\},
\]
whereas for the embedded Floquet code the optimization is
performed at every time slice,
\begin{equation}
d_{\mathrm{emb}}
=
\min_t\min\{d_X(t),d_Z(t)\}.
\end{equation}

We use MILP to calculate the distances of the codes in Tables~\ref{tab:2bga-weight4-floquet} and~\ref{tab:floquet-amc4-tableI}.

\subsection{Logical error rates under EM3 noise}
\begin{figure}[t]
\centering
\includegraphics{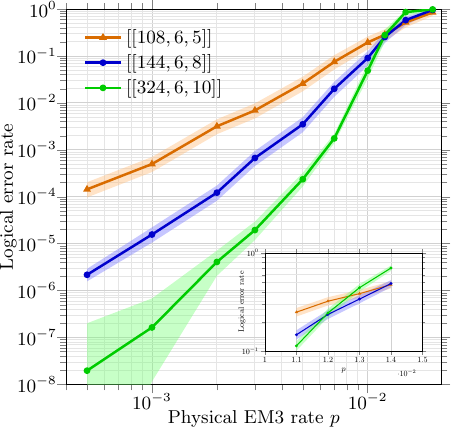}
\caption{Logical failure probability for the $[[108,6,5]]$, $[[144,6,8]]$, and $[[324,6,10]]$ Floquet AMC4 codes under EM3 noise, decoded with \texttt{tesseract-long-beam}. Each Stim circuit tracks six $Z$-basis memory observables. The simulations use $r_{\mathrm{mem}}=d_{\rm emb}$ noisy local rounds. The inset shows a magnified view of the crossing region.}
\label{fig:logical-rate}
\end{figure}

We simulate the logical error rates of three codes from Table~\ref{tab:floquet-amc4-tableI}: $[[108,6,5]]$, $[[144,6,8]]$, and $[[324,6,10]]$. The local Floquet round has period \(T_{\mathrm{local}}=6\), and we consider a memory experiment with EM3 noise applied for $r_{\mathrm{mem}}$ rounds; see Fig.~\ref{fig:logical-rate}. Each simulation uses one clean boundary local round, $r_{\mathrm{mem}}$ noisy local rounds, and one final clean boundary local round. Thus, the full circuit contains $(r_{\mathrm{mem}}+2)T_{\rm local}$ integer layers measured in units of rotated time, with noise applied only during the middle block. We choose $r_{\mathrm{mem}}=d_{\rm emb}$.

In our Floquet codes, each pairwise measurement is an $XX$ or $ZZ$ measurement on two physical worldline qubits. It is therefore natural to use the native two-body measurement noise model EM3~\cite{Gidney2021,HiggottBreuckmann2023,Fahimniya2025}.  In the EM3 model, a noisy two-qubit Pauli-product measurement
\(M_{PP}(p)\) measures the Pauli product and, with probability \(p\),
samples
\[
    \{I,X,Y,Z\}^{\otimes 2}\times\{\mathrm{no\ flip},\mathrm{flip}\}
\]
uniformly.  The component \((I,I,\mathrm{no\ flip})\) is the trivial identity component of this mixing channel.  
The flip component reports the wrong measurement result, while the Pauli component is applied to the two measured qubits after the measurement~\cite{Gidney2021}.  We note that this is a measurement-native circuit noise model.

In the Stim circuit, the reported-result flip is implemented by a noiseless flag record.  Each abstract measurement record $m_j$ is represented by two Stim measurement records:
\(
    m_j^{\rm eff} = m_j^{\rm MPP} \oplus m_j^{\rm flag}.
\)
The EM3 channel is implemented using five binary fault components:
$X_{q_1}$, $Z_{q_1}$, $X_{q_2}$, $Z_{q_2}$, and $\mathrm{flip}$. To reproduce the EM3 channel,  
the independent-component probability $p_{\rm ind}$ should satisfy $2p_{\rm ind}=1-(1-p)^{1/2^{N_e-1}}$, where $N_e=5$ is the number of basis errors~\cite{Chao2020}.
All nonzero products of these five components are inserted as correlated-error mechanisms after the ideal two-qubit measurement.  This reproduces the uniform 32-component EM3 channel.  

In our simulations, we use a local Pauli-web detector set.  
We represent each pairwise measurement by a binary symplectic row
\begin{equation}
    h_e=(x_e\mid z_e)\in\F^{2n_{\rm floq}}.
\end{equation}
Stacking all rows gives
\begin{equation}
\label{eq:Dpair}
    {\mathcal M}=(\Hx\mid\Hz).
\end{equation}
Because we are dealing with CSS codes, $\Hx$ and $\Hz$ have disjoint support. The compact $X$ and $Z$ detectors are found by considering $\ker(\Hx[\R]^T)$ and $\ker(\Hz[\R]^T)$, respectively. Here $\Hx[\R]$ and $\Hz[\R]$ are the row-restricted submatrices indexed by $\R$. We minimize the number of measurements by restricting the calculation to a compact row set $\R$ corresponding to four time layers. We find three compact detector templates seeded on a $Z$ half-cell and three seeded on an $X$ half-cell. All other detectors are obtained by translating these templates.
By construction, for a detector template $\Gamma$ we have
\(
    \bigoplus_{j\in \Gamma} m_j^{\rm eff}=0
\)~\cite{Gidney2021,Gidney2021stim}.

For the CSS memory experiments reported here, we select a complete same-basis set of $k=6$ independent $Z$-type observables. The plotted error count is the total block failure probability: a shot is counted as an error if the decoder predicts any of the six observable bits incorrectly.
The Stim circuits are sampled and decoded with Sinter.  The decoder is \texttt{tesseract-long-beam}, using the standard Tesseract Sinter registry~\cite{sinter,tesseractdecoder,YeWeckerDelfosse2026}.
Figure~\ref{fig:logical-rate} shows the logical error rate for a memory experiment for $r_{\rm mem}=d_{\rm emb}$.
We observe strong performance and efficient error suppression compared with other families of Floquet codes~\cite{Stairway2026,HiggottBreuckmann2023}. In particular, the $[[144,6,8]]$ code shows a level of error suppression at $p=10^{-3}$ similar to that of the $[[288,14,10]]$ Stairway code~\cite{Stairway2026}, while outperforming it at larger error rates. We also observe a clear advantage over the $[[256,10,4]]$ semi-hyperbolic code~\cite{HiggottBreuckmann2023}, despite having a similar encoding rate.

To estimate the pseudothreshold, we study three codes from the same family, $[[108,6,5]]$, $[[144,6,8]]$, and $[[324,6,10]]$, all of which are locally equivalent to four-dimensional toric codes. From the curve crossing in Fig.~\ref{fig:logical-rate}, we estimate the pseudothreshold to be approximately $1.2\%$. This value is close to that obtained for static AMC4 codes decoded under a circuit-level noise model~\cite{LinAMC2026}.
In Table \ref{tab:floquet-amc4-tableI}, we also present the Bravyi--Poulin--Terhal (BPT) metric for constructed AMC codes $k d^2_{\rm emb}/n$~\cite{BravyiPoulinTerhal2010}. 
The constructed AMC4 codes attain larger finite-size values
of the BPT metric $kd^2_{\rm emb}/n$ than the standard honeycomb code,
for which $kd^2_{\rm emb}/n=1/3$. 
Hyperbolic lattices can support homological models with an
extensive homology rank and a nonzero limiting
rate~\cite{JiangDumerKovalevPryadko2019}.
For finite-rate hyperbolic code
families, the logarithmic distance scaling implies the
asymptotic bound
\(
kd^2_{\rm emb}/n=O(\log^2 n)
\)
~\cite{Delfosse2013Tradeoffs,HiggottBreuckmann2024,Fahimniya2025}.

\section{Conclusions}
\label{sec:conclusions}

We have introduced Floquet Abelian multicycle codes, a
measurement-driven extension of AMC codes that combines
the multiblock homological structure of group-algebra chain
complexes with a periodic sequence of native two-qubit
Pauli measurements. As part of this construction, we
discussed a lattice description of Abelian
two-block group-algebra codes~\cite{WangPryadko2022,Arnault2026}. We
then extended this description to a general level-$j$,
$D$-dimensional AMC complex. The resulting quotient-lattice
unit cell is determined directly by the row- and column-block
structure of the two CSS check matrices, and the construction
applies to arbitrary finite Abelian group algebras.

The static quotient lattice was lifted to spacetime by
introducing a rotated-time covector and a compatible
periodicity lattice $\Lambda'$. When the check and data
spiders have even valence and admit a time-oriented port
matching, the edges of the foliated ZX network can be
reinterpreted as physical-qubit worldlines. Each many-legged
spider can subsequently be decomposed into a sequence of
two-qubit $XX$ or $ZZ$ measurements. A Floquet AMC code is
therefore specified not only by its underlying AMC complex
but also by the spacetime periodicity lattice and the local
worldline matching. These additional choices determine the
number of physical qubits, the global worldline period, the
instantaneous stabilizer group, and the embedded distance.

We applied this construction to generalized bicycle codes
and level-$2$ AMC4 codes built from weight-two
group-algebra elements. For the AMC4 matching considered
here, the worldlines have a local period of six rotated-time
layers, with each layer comprising four measurement
subrounds. We confirmed that the AMC4 examples considered
here retain six protected logical qubits and calculated their
embedded distances using mixed-integer linear programming.
The resulting finite-size codes exhibit competitive values of
the Bravyi--Poulin--Terhal metric and provide compact
examples of Floquet memories derived from codes locally
equivalent to four-dimensional toric codes.

We also constructed local Pauli-web detector templates and
simulated the $[[108,6,5]]$, $[[144,6,8]]$, and
$[[324,6,10]]$ Floquet AMC4 circuits under the
measurement-native EM3 noise model. With the detector and
decoder implementation used here, the logical failure
probability is strongly suppressed as the code distance
increases for physical error rates below the crossing region,
and the three curves yield an estimated pseudothreshold of
approximately $1.2\%$. In particular, the $[[144,6,8]]$
code provides error suppression comparable to that of larger
high-rate Floquet constructions while using a smaller
physical register. These results suggest that the redundancy and
confinement inherited from the AMC chain complex can be
converted into a useful spacetime detector structure without
requiring direct measurements of the original weight-six
stabilizer generators.

A substantial design space remains unexplored. In
particular, we did not consider any Floquet AMC3 examples.
A systematic search over $\Lambda'$, local port matchings,
and group-algebra generators may yield codes with lower
physical-qubit overhead, shorter global periods, and larger
embedded or circuit-level distances. Future realizations
based on noncyclic Abelian groups, nonuniform generator
weights, and higher-dimensional AMC complexes may clarify
how much of the single-shot and few-shot behavior of static
AMC codes can be retained in a Floquet setting. Floquet AMC
codes therefore provide a general framework linking compact
group-algebra qLDPC codes, higher-dimensional homological
redundancy, and measurement-native quantum error
correction, and may facilitate the construction of efficient
quantum memories.

\begin{acknowledgments} We gratefully acknowledge useful discussions with L.~Pryadko. This work was supported by the U.S. Department of Energy, Office of
Science, Basic Energy Sciences, under Award No. DE-SC0021019.  This work used
the Holland Computing Center of the University of Nebraska, which receives support from the UNL Office of
Research and Innovation, and the Nebraska Research Initiative.
\end{acknowledgments} 

\bibliographystyle{apsrev4-2}
\bibliography{amc}

\end{document}